\documentclass{ws-procs975x65}

\newcommand{\beeq}{\begin{equation}}
\newcommand{\eneq}{\end{equation}}
\newcommand{\beqn}{\begin{eqnarray}}
\newcommand{\eeqn}{\end{eqnarray}}

\def\la{\raise.16ex\hbox{$\langle$}\lower.16ex\hbox{}  }
\def\ra{\, \raise.16ex\hbox{$\rangle$}\lower.16ex\hbox{} }
\def\ep{\epsilon}
\def\psibar{ \psi \kern-.65em\raise.6em\hbox{$-$} \lower.6em\hbox{} }

\def\E{\mathcal{E}}

\def\R{\mathcal{R}}
\def\onehalf{\hbox{$\frac{1}{2}$}}
\def\pl{{\rm pl}}

\begin{document}

\title{False Vacuum Lumps with the Fermionic Core}

\author{Yutaka Hosotani\footnote{\uppercase{W}ork partially
supported by grants 13135215 and 13640284 of the \uppercase{M}inistry of 
\uppercase{E}ducation and \uppercase{S}cience of \uppercase{J}apan.}}

\address{Department of Physics, Osaka University,
Toyonaka,  Osaka 560-0043, Japan\\ 
E-mail: hosotani@phys.sci.osaka-u.ac.jp}

\author{Ramin G.\ Daghigh}

\address{Department of Chemistry and Physics, 
Arkansas State University\\ 
P.O.\ Box 419, State University, AR 72467-0419, USA\\
E-mail: rdaghigh@astate.edu}  


\maketitle

\abstracts{
Stable gravitating lumps with a false vacuum core surrounded 
by the true vacuum in a scalar field potential exist in the presence of
fermions at the core.   
These objects may exist in the universe at various scales.  
}

\section{Scalar field lumps with fermions}

When a scalar field potential has two non-degenerate minima,  the absolute
minimum of the potential corresponds to the true vacuum, while the other 
to the false vacuum.  
If the entire universe is in a false vacuum, it decays into
the true vacuum through bubble creation by quantum tunneling.
If the size of a false vacuum lump is smaller than the critical radius, the lump
would quickly decay, with its energy dissipating to the spatial infinity.
If its size is larger than the critical radius, the 
lump becomes a black 
hole.\cite{Daghigh2}${}^,$\cite{Hosotani2}${}^,$\cite{NoGo}

In this report we  demonstrate that 
a static false vacuum lump becomes stable if there are
additional fermions coupled to the scalar field.\cite{Daghigh1}
Yukawa interactions play a key role
in making such a structure possible. 
We stress that  these gravitating lumps
are  quite different from  boson stars,\cite{BosonStar} 
 Q-balls\cite{Qball} and Fermi-balls.\cite{Fball} 

We consider a system consisting of a scalar field $\phi$ and a fermion field
$\psi$ in Einstein gravity.  Its Lagrangian density is given by
\beeq
\mathcal{L}=\frac{1}{16\pi G} \R +\frac{1}{2}g^{\mu\nu}\partial_\mu\phi
\partial_\nu\phi-V[\phi]- (g \phi + m_0) \psibar \psi + \cdots
\label{model1}
\eneq
where $\R$ and  $V[\phi]$  are the scalar curvature and the scalar 
potential, respectively.   We take a potential
\beeq
V[\phi] \, = \frac{\lambda}{4} (\phi-f_2)
\bigg\{ \phi^3-\frac{1}{3}(f_2+4f_1)\phi^2
  - \frac{1}{3}f_2(f_2-2f_1)(\phi+f_2) \bigg\}  ~, 
\label{potential1}
\eneq
which satisfies $V'[\phi] = \lambda\phi(\phi-f_1)(\phi-f_2)$ and
$V[f_2] = 0$.

In a sherically symmetric configuration  (fig.\ 1), 
$g \la \psibar \psi \ra =  \rho_0 \theta (R_1 -r)$ and
\beeq
ds^2 =\frac{H(r)}{p(r)^2}dt^2 -\frac{dr^2}{H(r)}
  -r^2(d\theta^2+\sin^2\theta  d\varphi^2) ~~.
\eneq
The Einstein equations and the scalar field equation reduce to
\beqn
&&\hskip -1cm
\phi''(r)+\bigg( \frac{2}{r}+4\pi Gr 
   {\phi'(r)}^2+\frac{H'}{H} \bigg) \phi'(r)
=\frac{1}{H}\Big (V'[\phi]+ g \la \psibar\psi \ra  \Big) ~~,  \cr
&&\hskip -1cm
H = 1 - {2G\over r} \int_0^r 4\pi r^2 dr \, T_{00}[\phi] ~~.
\label{scalar2} 
\eeqn
We look for a solution in which $\phi \sim f_1$ inside the lump,
whereas $\phi \sim f_2$ outside.

\begin{figure}[t]
\centering 
\leavevmode 
\mbox{
\epsfxsize=5.cm 
\epsfbox{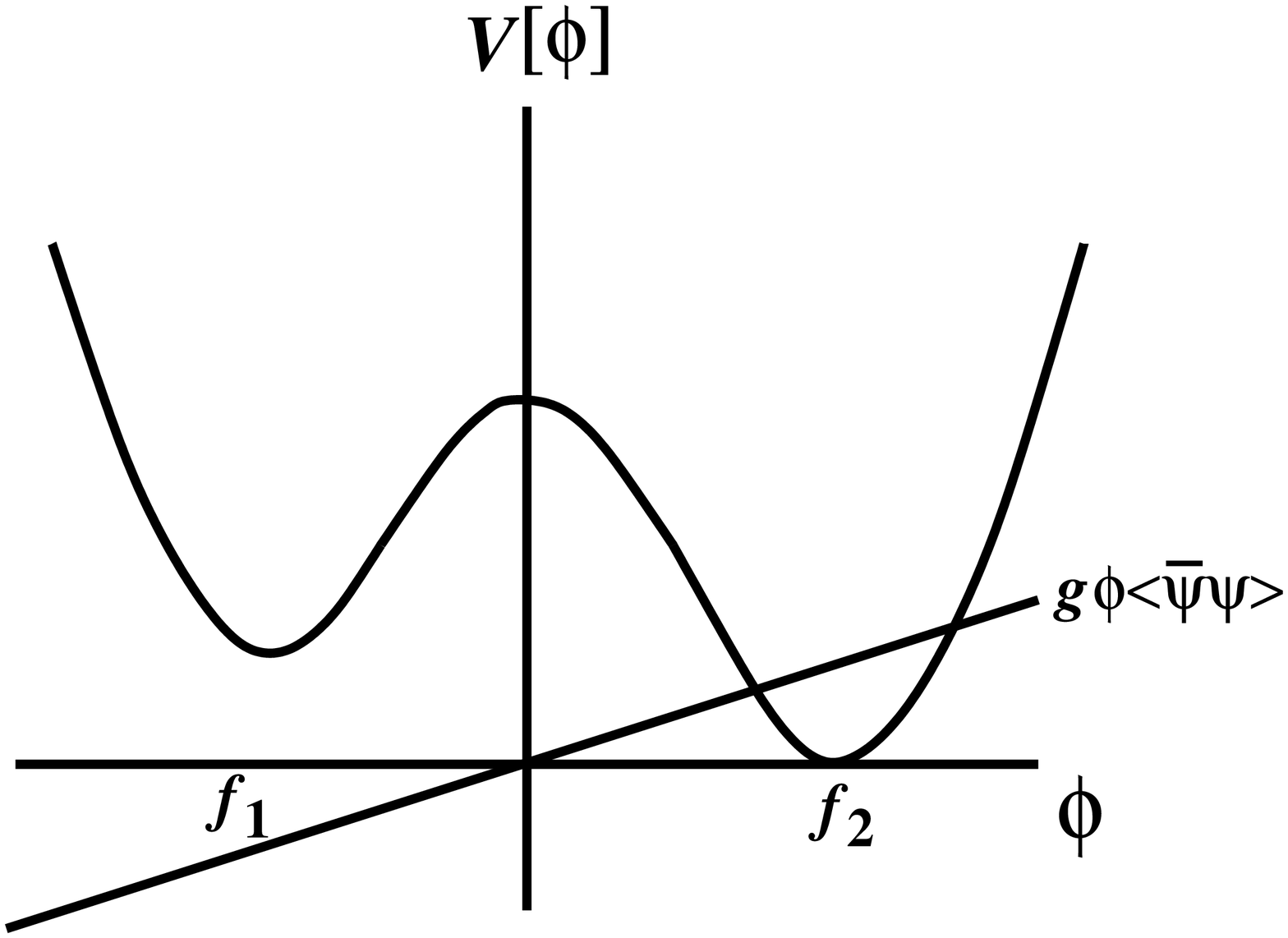}}
\hskip .5cm 
\mbox{
\epsfxsize=5.cm 
\epsfbox{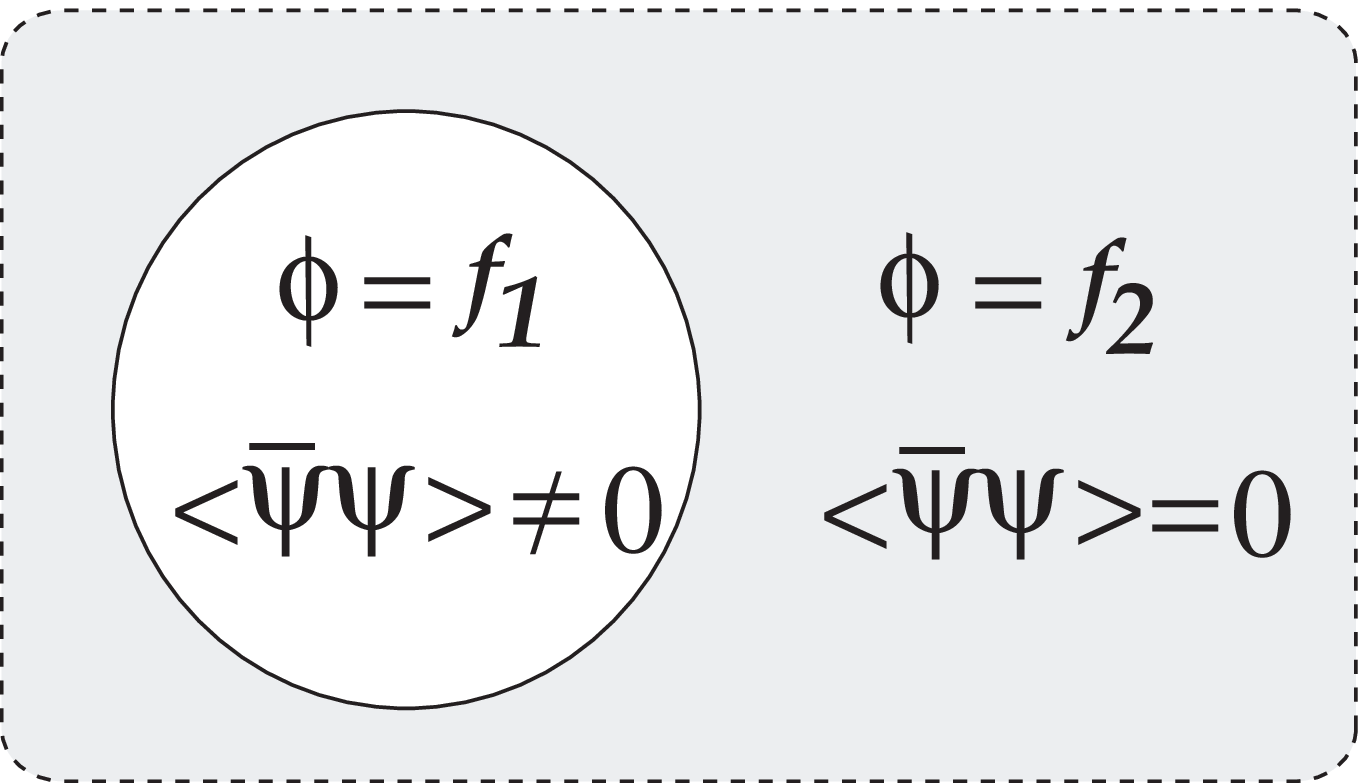}}
\caption{The scalar field potential and the false vacuum lump.}
\end{figure}

\section{Solutions}

Fermions  have a mass $|m_0 +f_S| \equiv m$ inside  a lump.  We consider the 
case in which $m \ll m_0 + f_2$ and the fermion gas inside the lump is
nonrelativistic so that $\la \psibar\psi \ra \sim \la \psi^\dagger \psi \ra
= \rho_n$.  Let $R$ be the radius of the lump.  The total fermion number 
$(4\pi R^3/3) \rho_n $ is kept fixed.  The total energy of the lump is
$E (R) \sim \big\{ {\E}_f + \ep (\rho_n) \big\} 
   \,  (4\pi R^3/3) + 4\pi R^2 \sigma$ where
$\E_f = m \rho_n + 3 (3\pi^2)^{2/3} \rho_n^{5/3} / 10 m$. 
The minimum of
$U[\phi] = V[\phi] + g \rho_n \phi$ is located at $\phi=f_S$, i.e.\ $U'[f_S]=0$.
Then $\ep \equiv U[f_S] \sim \ep_0 = V[f_1]$ for small $\rho_n$,
while $\ep \sim - (f\rho_n)^{4/3}$ for large $\rho_n$.  $\sigma$ is  the surface
tension resulting from varying $\phi$ in the boundary wall region.
$E (R)$ is minimized at $R=\bar R$ which gives the size of the lump
configuration.

\begin{figure}[t]
\centering 
\leavevmode 
\mbox{
\epsfxsize=5.cm 
\epsfbox{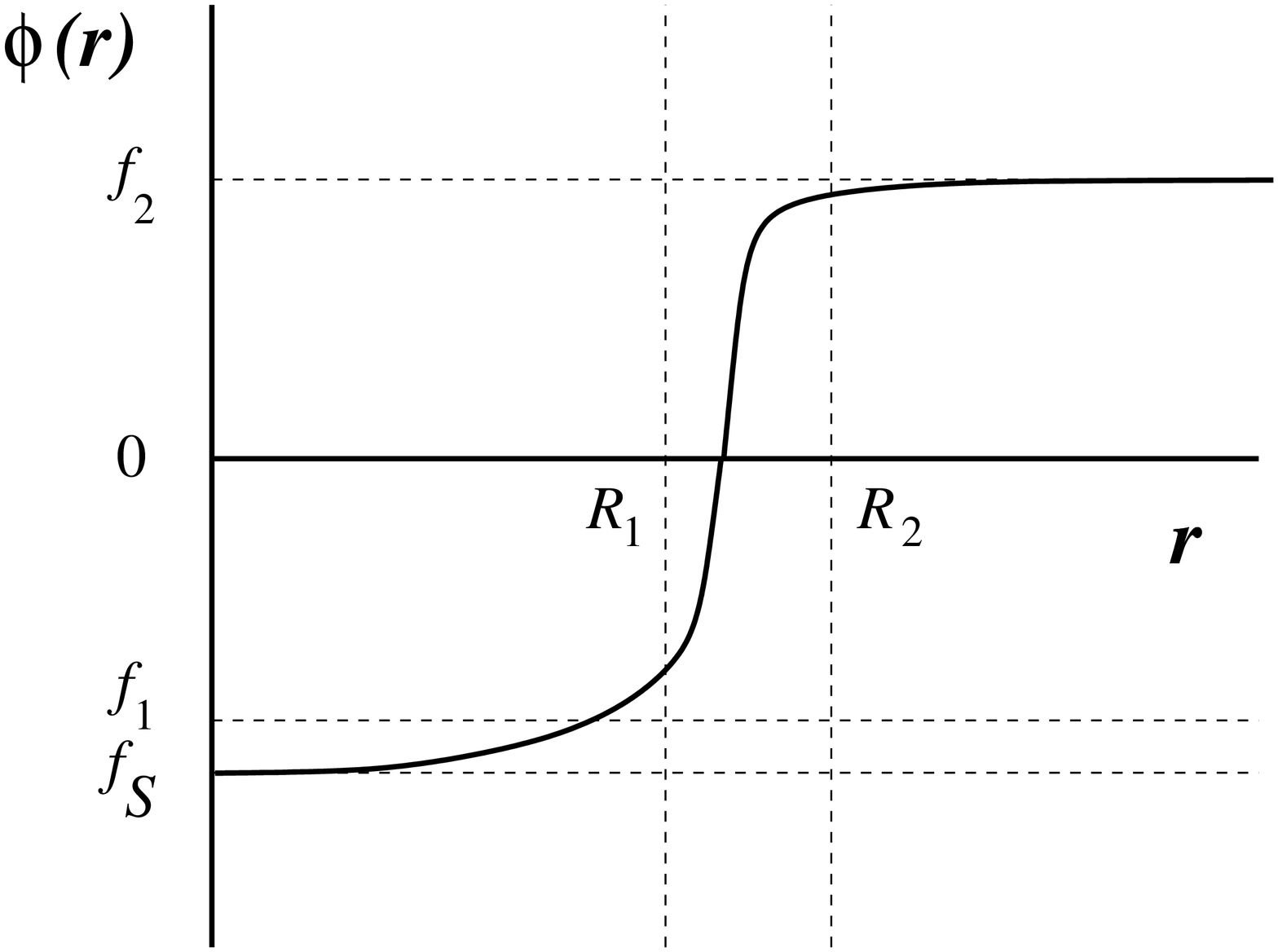}}
\hskip .5cm 
\mbox{
\epsfxsize=5.cm 
\epsfbox{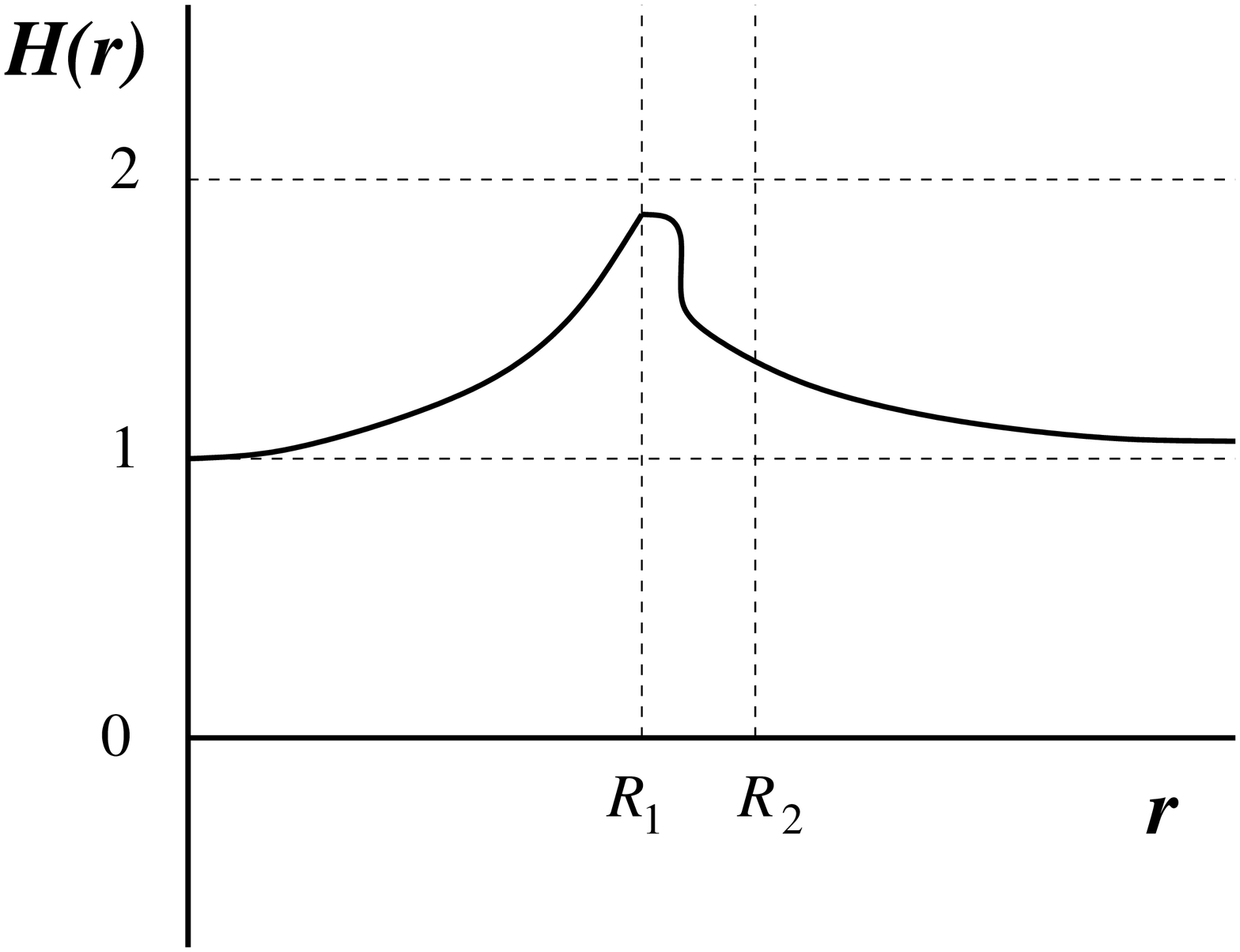}}
\caption{The behavior of $\phi(r)$ and $H(r)$ is shown schematically.
$R_2 - R_1 \ll R_1$}
\end{figure}

The behavior of the solution is displayed in fig.\ 2.
$\phi(r)$ makes a sharp transition from $f_S$ to $f_2$ in $R_1 < r < R_2$.  
The wall  is very thin; $R_2 - R_1 \sim 1/ \sqrt{\lambda} f \ll R_1$ where
$f=(|f_1|+f_2)/2$.  The geometry is approximately 
anti-de Sitter inside the lump ($\ep < 0$) and Schwarzschild outside the lump.

Inside the lump
\beqn
&&\hskip -1cm
\phi(r) = f_S + \delta \phi (0) \cdot
F \Big( \frac{3}{4} + \kappa , \frac{3}{4} -  \kappa ,
  \frac{3}{2} ; -\frac{r^2}{a_f^2} \Big) ~~, \cr
&&\hskip -1cm
a_f = \sqrt{ {-3\over 8\pi G \ep}} ~~,~~
\kappa = \onehalf \sqrt{  a_f^2 V''(f_S)  + \hbox{${9\over 4}$} } ~~.
\label{phi1}
\eeqn
$F$ is Gauss' hypergeometric function.  Outside the lump
$\phi \sim f_2$ and $H = 1 - (2G \tilde M/r)$.  For $R_1 < r < R_2$
the solution is determined  numerically. 

Given appropriate values for $\rho_n$ and $R_1$, a solution is found
by choosing $\delta \phi(0)$.  In the numerical investigation
$\delta\phi(R_1)$, instead of $\delta \phi(0)$, is fine tuned to obtain a
solution.   For instance, with input parameters $\lambda=g=1$, 
$f/M_\pl = (f_2-|f_1|)/f =2 \cdot 10^{-4}$, $(f_1 - f_S)/f = 5 \cdot 10^{-3}$,
one finds that $a_f/l_\pl = 8.65 \cdot 10^7$, $- \ep/\ep_0 = 74.7$, 
$\kappa=12327$ and $\rho_n l_\pl^3 = 8.06 \cdot 10^{-14}$.  
For $R_1/l_\pl = 8 \cdot 10^7$, a solution is found with 
$\delta\phi( R_1)/f = 2.77 \cdot 10^{-2}$.  Using (\ref{phi1}),
one finds $\delta\phi(0)/f = 4.7 \cdot 10^{-8855}$ !  

We stress that solutions continue to exist for much smaller values of $f$.
False vacuum lump solutions are typically macroscopic  ($R_1 \gg f^{-1}$).
They may appear at various scales in the universe.  It is remarkable that
such objects appear in a very simple model consisting of a scalar field
and fermions.

%
%
%
%


\begin{thebibliography}{0}

\def\jnl#1#2#3#4{{#1}{\bf #2}~ #3 (#4)}

\def\PTP{{\em Prog.\ Theoret.\ Phys.\  }}
\def\PRD{{\em Phys.\ Rev.} D}
\def\CQG{\em Class.\ Quant.\ Grav. }
\def\NPB{{\em Nucl.\ Phys.} B}
\def\PLB{{\em Phys.\ Lett.} B}
\def\PR{{\em Phys.\ Rev.} }
\def\PRe{{\em Phys.\ Rep.} }





\bibitem{Daghigh2}
R.G.\ Daghigh, J.I.\ Kapusta, and Y.\ Hosotani, gr-qc/0008006.

\bibitem{Hosotani2}
Y. Hosotani, Soryushiron Kenkyu 
{\bf 103}  E91  (2001),   hep-th/0104006.

\bibitem{NoGo}
D.V.\ Gal'tsov  and J.P.S.\ Lemos,
\jnl{\CQG}{18}{1715}{2001},  gr-qc/0008076;

K.A.\ Bronnikov, \jnl{\PRD}{64}{064013}{2001},  gr-qc/0104092;

K.A.\ Bronnikov and G.N.\ Shikin, 
\jnl{\em Grav. Cosmol. }{8}{107}{2002}, gr-qc/0109027.


\bibitem{Daghigh1}
R.G.\ Daghigh and Y.\ Hosotani, 
\jnl{\PTP}{110}{1151}{2003},  gr-qc/0307075.




\bibitem{BosonStar}
D.J.\ Kaup, \jnl{\PR}{172}{1331}{1968};

E.W.\ Mielke and R.\ Scherzer, \jnl{\PRD}{24}{2111}{1981};

P.\ Jetzer, \jnl{\PRe}{220}{163}{1992}.

\bibitem{Qball}
S. Coleman, \jnl{\NPB}{262}{263}{1985};

A. Kusenko,  \jnl{\PLB}{404}{285}{1997}.
 
\bibitem{Fball}
A. L. Macpherson and B.A.\ Campbell, \jnl{\PLB}{347}{205}{1995};

J.R. Morris, \jnl{\PRD}{59}{023513}{1998};

T. Yoshida, K. Ogure, and J. Arafune, \jnl{\PRD}{68}{023519}{2003}.




\end{thebibliography}
\end{document}